\documentclass[12pt]{article}
\usepackage{latexsym}
\begin{document}
\begin{center}

{\bf Viscous Modified Gravity on a RS Brane Embedded in AdS$_5$}

\vspace{1cm} Iver Brevik\footnote{E-mail: iver.h.brevik@ntnu.no}

\bigskip

Department of Energy and Process Engineering, Norwegian University
of Science and Technology, N-7491 Trondheim, Norway

\bigskip

\today
\end{center}

\begin{abstract}

We consider a modified gravity fluid on a Randall-Sundrum II brane
situated at $y=0$, the action containing a power $\alpha$ of the
scalar curvature. As is known from 4D spatially flat modified
gravity, the presence of a bulk viscosity may drive the cosmic
fluid into the phantom region ($w<-1$) and thereafter inevitably
into the Big Rip singularity, even it is initially nonviscous and
lies in the quintessence region ($w>-1$). The condition for this
to occur is that the bulk viscosity contains the power
$(2\alpha-1)$ of the scalar expansion. We combine this with the 5D
RS II model, and find that the Big Rip, occurring for $\alpha
>1/2$, carries over to the metric for the bulk metric, $|y|>0$.
Actually, the scale factors on the brane and in the bulk become
simply proportional to each other.

\end{abstract}


\section{Introduction}

Modified gravity theories in 4D continue to attract interest; this
obviously being related to observations, for instance the measured
redshifts from type Ia supernovae
\cite{reiss98,perlmutter99,tony03}. The data may be reconciled
with the concept of dark energy, with a cosmic fluid with a
complicated equation of state, or with a scalar field having
quintessence or phantom behavior. An extensive recent review is
given by Copeland {\it et al.} \cite{copeland06}. The equation of
state for the cosmic fluid is conventionally written as $p=w\rho$,
where $w=-1$ corresponds to a vacuum fluid (cosmological
constant), $-1<w<-1/3$ to a quintessence fluid, and $w<-1$ to a
phantom fluid, having the bizarre property of predicting a Big Rip
singularity in the future.

In the present paper we combine essentially two kinds of theories:

1)  First, we assume the modified 4D action in the form given by
Eq.~(\ref{18}) below. The integrand contains  a power $\alpha$ in
the scalar curvature. This model has recently been studied by
Abdalla {\it et al.} \cite{abdalla05}. As a generalization, we
include a bulk viscosity in the fluid being proportional to the
$(2\alpha-1)$'th power of the scalar expansion. Cf. Eq.~(\ref{31})
below. This viscous model has been studied repeatedly in the
recent past \cite{brevik05,brevik05a,brevik06,brevik06b}. A
striking property of this kind of theory is that it leads to the
Big Rip singularity, if $\alpha>1/2$.

2) The next step in our analysis is to combine the above 4D theory
with the 5D Randall-Sundrum II model, \cite{randall99}, where
there is a spatially flat brane situated at $y=0$, surrounded by
an AdS space. Our main result is to show how the mentioned Big Rip
singularity on the brane becomes transferred to the bulk: the
scale factor away from the brane ($|y|>0$) becomes actually
proportional to the scale factor on the brane itself. In this
sense the 4D and 5D gravity theories are  closely intertwined. In
this respect there is no essential difference between the modified
gravity and the Einstein gravity. They behave qualitatively in the
same way, only with the characteristic difference that the
strength of the singularity becomes larger for increasing values
of $\alpha$.

\section{Basic formalism}

Assume, as mentioned,  that there is a spatially flat ($k=0$)
brane  located at the fifth dimension $y=0$, surrounded by an
Anti-de Sitter (AdS) space. If the five-dimensional cosmological
constant $\Lambda (<0)$ is different from zero, this model is the
 Randall-Sundrum II model (RSII) \cite{randall99}. We
shall take the metric to have the form
\begin{equation}
ds^2= -n^2(t,y)dt^2+a^2(t,y)\delta_{ij}dx^idx^j +dy^2, \label{1}
\end{equation}
where $n(t,y)$ and $a(t,y)$ are determined from Einstein's
equations
\begin{equation}
R_{AB}-\frac{1}{2}g_{AB}R +g_{AB}\Lambda=\kappa^2T_{AB}. \label{2}
\end{equation}
The coordinate indices are numbered as $x^A=(t,x^1,x^2,x^3,y)$,
with $\kappa^2=8\pi G_5$ the five-dimensional gravitational
coupling. Einstein's equations in a coordinate basis with the
metric (\ref{1}) have been given before (cf., for instance,
Refs.~\cite{brevik04,binetruy00,binetruy00a,brevik02,brevik04a,brevik06a},
but for convenience we give them also here:
\begin{equation}
3\left\{ \left( \frac{\dot{a}}{a}\right)^2 -n^2
\left[\frac{a''}{a} +\left(\frac{a'}{a} \right)^2 \right]
\right\}-\Lambda n^2=\kappa^2 T_{tt}, \label{3}
\end{equation}
\begin{equation}
a^2\delta_{ij} \Bigg\{ \frac{a'}{a}\left( \frac{a'}{a}+\frac{2n'}{n}\right) +\frac{2a''}{a}+\frac{n''}{n}\nonumber\\
+\frac{1}{n^2}\Big[
\frac{\dot{a}}{a}\left(-\frac{\dot{a}}{a}+\frac{2\dot{n}}{n}\right)-\frac{2\ddot{a}}{a}\Big]
+ \Lambda \Bigg\} =\kappa^2 T_{ij}, \label{4}
\end{equation}
\begin{equation}
3\left( \frac{\dot{a}}{a}\frac{n'}{n}-\frac{\dot{a}'}{a}
\right)=\kappa^2T_{ty}, \label{5}
\end{equation}
\begin{equation}
3\Bigg\{\frac{a'}{a}\left(\frac{a'}{a}+\frac{n'}{n}\right)-\frac{1}{n^2}\Big[\frac{\dot{a}}{a}\left(
\frac{\dot{a}}{a}-\frac{\dot{n}}{n}\right)
+\frac{\ddot{a}}{a}\Big]\Bigg\}+\Lambda=\kappa^2 T_{yy}. \label{6}
\end{equation}
Overdots and primes mean derivatives with respect to $t$ and $y$,
respectively. As the $5D$ space outside the brane is taken to be
empty, the components of $T_{AB}$ are different from zero only on
 the brane.

Consider next the form of $T_{AB}$. Let $U^\mu=(U^0, U^i)$ (Greek
indices $\mu,\nu \in [0,3]$) be the fluid's four-velocity on the
brane, and let $\sigma$ denote the  brane tension, assumed
constant.  Moreover, let $h_{\mu\nu}=g_{\mu\nu}+U_\mu U_\nu$ be
the projection tensor, and $\tilde{p}=p-3H_0\zeta$ the effective
pressure with $H_0=\dot{a}_0/a_0$ the Hubble parameter on the
brane and $\zeta$ the bulk viscosity. The shear viscosity is
omitted due to the assumed spatial isotropy.

It might be noted here that a viscous fluid may be considered as a
specific example of an inhomogeneous equation-of-state fluid
introduced in Refs.~\cite{nojiri05a,capozziello06}.

As gauge condition we take $n_0(t)=1$, this meaning that the
proper time on the brane is the same as the cosmological time. The
energy-momentum tensor can accordingly be written as
\begin{equation}
T_{AB}=\delta(y)(-\sigma g_{\mu\nu}+\rho U_{\mu}U_\nu +\tilde{p}\,
h_{\mu\nu})\delta_A^\mu\,\delta_B^\nu. \label{7}
\end{equation}
We shall work in the orthonormal frame where $U^\mu=(1,0,0,0)$,
and let a subscript zero refer to the brane.

The junction conditions at $y=0$ have now to be taken into
account. They express that the metric is continuous across the
brane, but its derivatives are not. From Eqs.~(\ref{3}) and
(\ref{4}) we get, for the distributional parts,
\begin{equation}
\frac{[a']}{a_0}=-\frac{1}{3}\kappa^2(\sigma+\rho), \label{8}
\end{equation}
\begin{equation}
[n']=\frac{1}{3}\kappa^2(-\sigma+2\rho+3\tilde{p}), \label{9}
\end{equation}
where $[a']=a'(0^+) -a'(0^-)$, and similarly for $[n']$. For the
nondistributional parts we get
\begin{equation}
\left(
\frac{\dot{a}}{na}\right)^2-\frac{a''}{a}-\left(\frac{a'}{a}\right)^2=\frac{1}{3}\Lambda,
\label{10}
\end{equation}
\[ \frac{a'}{a}\left(
\frac{a'}{a}+\frac{2n'}{n}\right)+\frac{2a''}{a}+\frac{n''}{n} \]
\begin{equation}
+\frac{1}{n^2}\left[
\frac{\dot{a}}{a}\left(-\frac{\dot{a}}{a}+\frac{2\dot{n}}{n}\right)-\frac{2\ddot{a}}{a}\right]=-\Lambda.
\label{11}
\end{equation}
Assuming no energy flux to occur from the brane, we have
$T_{ty}=0$. It implies that
\begin{equation}
n(t,y)=\frac{\dot{a}(t,y)}{\dot{a}_0(t)} \label{12}
 \end{equation}
for arbitrary $y$. Then from Eq.~(\ref{10}) we get, upon
integration with respect to $y$,
\begin{equation}
\left(\frac{\dot{a}}{na}\right)^2=\frac{1}{6}\Lambda+\left(\frac{a'}{a}\right)^2+\frac{C}{a^4}.\label{13}
\end{equation}
Here $C=C(t)$ is an integration constant with respect to $y$. The
$C$ term is called the "radiation term"; as it is not of main
interest here, it will be omitted in the following.

Now setting $y=0$ we get for $H_0=\dot{a}_0/a_0$
\begin{equation}
H_0^2=\frac{1}{6}\Lambda+\frac{\kappa^4}{36}(\sigma+\rho)^2.
\label{14}
\end{equation}
Recall that $\Lambda$ and $\sigma$ are constants, while
$\rho=\rho(t)$. The generalized Friedmann equation (\ref{14}) can
be contrasted with the conventional Friedmann equation in
four-dimensional space (still with $k=0$),
\begin{equation}
H_0^2=\frac{1}{3}\Lambda_4+\frac{1}{3}\kappa_4^2 \,\rho,
\label{15}
\end{equation}
with $\kappa_4^2=8\pi G_4$. The essential new feature of
Eq.~(\ref{14}) is thus the occurrence of a $\rho^2$ term.

Let us observe the solution for $a_0(t)$ from Eq.~(\ref{14}) if
$\rho=0$:
\begin{equation}
a_0(t)\Big|_{\rho=0}=\frac{1}{2\sqrt{\lambda}}\exp[\sqrt{\lambda}(t+c_0)],
\label{16}
\end{equation}
where
\begin{equation}
\lambda=\frac{1}{6}\Lambda+\frac{1}{36}\kappa^4 \sigma^2,
\label{17}
\end{equation}
$c_0$ being a new integration constant. The scale factor is thus
exponentially increasing, qualitatively as in the ordinary de
Sitter case.

\section{Modified gravity on the brane}

In this section we consider the  fluid - Einstein or modified
fluid - on the brane $y=0$. We shall derive how the Hubble
parameter $H$ varies with time $t$, leading eventually to the Big
Rip.

We adopt the following 4D gravity model:
\begin{equation}
S=\frac{1}{2\kappa_4^2} \int d^4 x \sqrt{-g}\,(f_0R^\alpha +L_m),
\label{18}
\end{equation}
where $f_0$ and $\alpha$ are constants ($\alpha$ may in principle
be negative), and $L_m$ is the matter Lagrangian. This model has
been studied before; cf., for instance,
Refs.~\cite{abdalla05,brevik05a,brevik06}. The case $f_0=1,\,
\alpha=1$ corresponds to Einstein's gravity. (More complicated
$f(R)$ theories have been discussed at various places, for
instance in Refs.~\cite{nojiri05a,capozziello06,bronnikov07}. We
may also mention here that a general review of modified gravity
can be found in Ref.~\cite{nojiri07}, and recent reviews of viable
$f(R)$ gravity theories unifying dark energy, inflation, and dark
matter, can be found in Refs.~\cite{nojiri08,nojiri08a}.)

The equations of motion following from the action (\ref{18}) are
\[ -\frac{1}{2}f_0 g_{\mu\nu}R^{\alpha}+\alpha f_0 R_{\mu\nu}
R^{\alpha-1} \]
\begin{equation}
-\alpha f_0\nabla_\mu\nabla_\nu R^{\alpha-1}+\alpha f_0
\,g_{\mu\nu}\nabla^2 R^{\alpha-1}=\kappa_4^2T_{\mu\nu}, \label{19}
\end{equation}
where $T_{\mu\nu}$ corresponds to the term $L_m$ in the
Lagrangian. (The proposal of taking $f(R)$ on the brane was
considered also in Ref.~\cite{nojiri05c}.)

We take the equation of state for the fluid to have the
conventional form
\begin{equation}
p=w\rho \equiv (\gamma-1)\rho \label{20}
\end{equation}
(more complicated forms for the equation of state have recently
been investigated by \cite{nojiri05a,capozziello06,sussman08}). If
$w=-1$ or $p=-\rho$ we have a "vacuum fluid", with bizarre
thermodynamic properties such as possibly negative entropies
\cite{brevik04b}. As is known, cosmological observations indicate
that the present universe is accelerating. Moreover, based upon
the observed data it has been conjectured that $w$ is a varying
function of time. For instance, as discussed in
Ref.~\cite{vikman05}, $w$ might have been around 0 at redshift $z
\sim 1$ and may be slightly less than -1 today. Perhaps is $w$
even an oscillating function in time. Recent discussions on
possible forms of the equation of state are given, for instance,
in Refs.~\cite{brevik07} and \cite{brevik07a}. In view of these
circumstances the analysis of a possible crossing of the phantom
barrier $w=-1$, from the quintessence region $(-1<w<-1/3)$ into
the phantom region $w<-1$, is obviously of physical interest. It
ought to be noted that both quintessence and phantom fluids lead
to the inequality $\rho+3p \leq 0$, thus breaking the strong
energy condition. The Big Rip singularity has been discussed in
various papers; cf., for instance,
Refs.~\cite{caldwell03,mcinnes02,barrow04,nojiri05,nojiri04}.
Specifically, the future singularities in modified gravity were
considered also in Ref.~\cite{nojiri08c}.

Of main interest is the (00)-component of Eq.~(\ref{19}).
Observing that $ R_{00}=-3\ddot{a}/a,\, R=6(\dot{H}+2H^2)$, as
well as $T_{00}=\rho$, we obtain
\begin{equation}
 \frac{1}{2}f_0 R^\alpha-3\alpha f_0(\dot{H}+H^2)R^{\alpha-1}
+3\alpha (\alpha-1)f_0 HR^{\alpha-2}\dot{R}=\kappa_4^2\,\rho.
\label{21}
\end{equation}
An important property of Eq.~(\ref{21}) is that the covariant
divergence of the LHS is equal to zero \cite{koivisto05},
\begin{equation}
\nabla^\nu T_{\mu\nu}=0, \label{22}
\end{equation}
just as in  Einstein's gravity. Energy-momentum conservation is a
consequence of the field equations. This leads to the energy
conservation equation
\begin{equation}
\dot{\rho}+(\rho+p)3H=9\zeta H^2. \label{23}
\end{equation}
We now differentiate the expression (\ref{21}) with respect to
$t$, and insert $\dot{\rho}$ from Eq.~(\ref{23}). After some
calculation we  obtain
\[ \frac{3}{2}\gamma
f_0 R^\alpha +3\alpha f_0[2\dot{H}-3\gamma
(\dot{H}+H^2)]R^{\alpha-1} \]
\begin{equation}
+3\alpha(\alpha-1)f_0[(3\gamma-1)H\dot{R}+\ddot{R}]R^{\alpha-2}+3\alpha(\alpha-1)(\alpha-2)f_0\dot{R}^2R^{\alpha-3}=9\kappa_4^2\zeta
H. \label{24}
\end{equation}
Recalling that $R=6(\dot{H}+2H^2)$, we see that this equation  is
a complicated nonlinear differential equation for $H(t)$. It is
best discussed in terms of examples.  We are interested in
solutions that are related to the Big Rip. We shall look for
solutions having the form
\begin{equation}
H=\frac{H_*}{X}, \quad {\rm where} \quad X \equiv 1-BH_* \,t.
\label{25}
\end{equation}
Here $H_*$ is the Hubble parameter at present time $t=0$ (usually
called $H_0$ but we are reserving the  subscript zero mainly for
 brane entities), and $B$ is a nondimensional  constant.
For Big Rip to occur, $B$ has to be positive.

\subsection{Einstein's gravity fluid}

As mentioned above, this case corresponds to $f_0=1,\, \alpha =1$.
As for the bulk viscosity, we shall take $\zeta$ to be
proportional to the scalar expansion $\theta=3H$ through a
proportionality constant, here called $\tau_E$,
\begin{equation}
\zeta=\tau_E\theta=3\tau_E H. \label{26}
\end{equation}
This form is of particular physical interest. Namely, as shown in
Ref.~\cite{brevik05}, if $\tau_E$ is large enough to satisfy the
condition
\begin{equation}
\chi \equiv -\gamma+3\kappa_4^2\,\tau_E >0, \label{27}
\end{equation}
then the equations of motion lead to the Big Rip singularity in a
finite time $t$. Even if one starts with a state where the fluid
is nonviscous and lies in the quintessence region ($\gamma
>0$), the imposition of a sufficiently large bulk viscosity will
drive it into the phantom region and thereafter inevitably into
the Big Rip.

From the governing equations we now get
\begin{equation}
B=\frac{3}{2}\chi, \label{28}
\end{equation}
\begin{equation}
H_*=\sqrt{\frac{1}{3}\kappa_4^2\,\rho_*}, \label{29}
\end{equation}
\begin{equation}
\rho_E=\frac{\rho_*}{X^2}, \label{30}
\end{equation}
where $\rho_*$ is the $t=0$ value of the energy density.

\subsection{Modified gravity fluid}

Assume now that $f_0$ and $\alpha$ are arbitrary. Let the bulk
viscosity for the modified fluid be denoted by $\zeta_\alpha$. As
in Refs.~\cite{brevik06,brevik06b} we model $\zeta_\alpha$ by
setting it proportional to the $(2\alpha-1)$'th power of the
scalar expansion:
\begin{equation}
\zeta_\alpha=\tau_\alpha \theta^{2\alpha-1}=\tau_\alpha
(3H)^{2\alpha-1}. \label{31}
\end{equation}
The main reason for this assumption is that it fits well with the
governing equation for $H$ as well as with our previous assumption
(\ref{26}): the time-dependent factors in Eq.~(\ref{24})
automatically drop out, and we remain with the following equation
determining $B$:
\[
(B+2)^{\alpha-1}
\Big\{9(2-\alpha)\gamma+3[\alpha+3\gamma+\alpha(2\alpha-3)(3\gamma-1)]B
\]
\begin{equation}
+6\alpha(\alpha-1)(2\alpha-1)B^2\Big\}=\frac{18\kappa_4^2}{f_0}\left(\frac{3}{2}\right)^\alpha
\tau_\alpha. \label{32}
 \end{equation}
 This equation is complicated, and is best discussed in terms of
 examples. For instance, if $\alpha=2$ and $\gamma=0$ (the
 latter condition corresponding to a vacuum fluid), then
 Eq.~(\ref{32}) yields the following cubic equation ($\tau_\alpha
 \rightarrow \tau_2$):
 \begin{equation}
 B^3+2B^2=\frac{9\kappa_4^2\tau_2}{8f_0}. \label{33}
 \end{equation}
 There exists one single positive root of this equation, as long as  the
 RHS is positive. This root is caused by viscosity,
 and leads to the Big Rip.

 We note also the general equation for $B$ following directly from
 the energy conservation equation (\ref{23}) for the modified fluid,
\begin{equation}
\dot{\rho_\alpha}+(\rho_\alpha+p_\alpha)3H=9\zeta_\alpha H^2,
\label{34}
\end{equation}
namely
\begin{equation}
B=-\frac{3\gamma}{2\alpha}+\frac{3\tau_\alpha}{2\alpha}\frac{(3H_*)^{2\alpha}}{\rho_*},
\label{35}
\end{equation}
where we used
\begin{equation}
\zeta_\alpha=\tau_\alpha \left(\frac{3H_*}{X}\right)^{2\alpha-1},
  \quad \rho_\alpha=\frac{\rho_*}{X^{2\alpha}}. \label{36}
\end{equation}
For simplicity we have assumed the same initial conditions at
$t=0$ for the modified fluid as for the Einstein fluid, viz.
$\rho_{*\alpha}=\rho_{*E} \equiv \rho_*$, $H_{*\alpha}=H_{*E}
\equiv H_*$.

\section{Implications for the 5D theory}

We are now equipped with the necessary background to see how the
modified fluid on the brane effects the 5D brane physics. Consider
first Eq.~(\ref{14}) on the brane (recall that this is a 5D, not a
4D,  equation). It is natural from a physical point of view to use
the expressions for $\rho(t)$ from the previous section as input
quantities in this equation. Comparison between Eqs.~(\ref{36})
and (\ref{30}) shows that we can regard
$\rho_\alpha=\rho_*/X^{2\alpha}$ as a generic equation common for
the two cases, only with $\alpha=1$ in the Einstein case. For the
5D scale factor $a_0(t)$ on the brane we obtain thus
\begin{equation}
H_0^2=\frac{1}{6}\Lambda+\frac{\kappa^4}{36}\left[\sigma+\frac{\rho_*}{(1-BH_*\,t)^{2\alpha}}\right]^2.
\label{37}
\end{equation}
As we shall be mainly interested in the behavior near Big Rip, we
consider times close to the singularity time $t_s=1/(BH_*)$, where
we get approximatively
\begin{equation}
\frac{\dot{a}_0}{a_0}=\frac{\kappa^2}{6}\frac{\rho_*}{(1-BH_*\,t)^{2\alpha}}.
\label{38}
\end{equation}
The quantities $\Lambda$ and $\sigma$, characteristic for 5D
theory, are here neglected. The solution of this equation is of
the form
\begin{equation}
a_0(t) \sim \exp \left[
\frac{(\kappa^2/6)\rho_*}{(2\alpha-1)(BH_*)^{2\alpha}(t_s-t)^{2\alpha-1}}\right].
\label{39}
\end{equation}
The scale factor on the brane has thus an essential singularity at
$t=t_s$, if $\alpha >1/2$. This behavior incorporates both the
Einstein gravity, and the $R^2$-modified gravity ($\alpha=2$),
considered in the previous section. The singularity is stronger
the larger is the value of $\alpha$. Moreover, the divergence
 is stronger than the power divergences found
for viscous 4D cosmology with account of quantum effects
\cite{brevik08}. If $\alpha<1/2$, $a_0$ does not diverge at $t_s$.
Note that there is a relationship between $\rho_*$ and $H_*$ in
the expression (\ref{39}) as following from Eq.~(\ref{14}) taken
at $t=0$,
\begin{equation}
H_*^2=\frac{1}{6}\Lambda+\frac{\kappa^4}{36}(\sigma+\rho_*)^2.
\label{40}
\end{equation}
Now return to the bulk case, considering Eq.~(\ref{13}) for
arbitrary $y$. When $C=0$ as assumed (recall that also $k=0$), we
obtain as AdS solution
\[
 a^2(t,y)=\frac{1}{2}a_0^2(t) \Bigg[ \left( 1+\frac{\kappa^4 \sigma^2}{6\Lambda}\right)
+ \left(1-\frac{\kappa^4 \sigma^2}{6\Lambda}\right) \cosh
(2\mu\,y)  \]
\begin{equation}
-\frac{\kappa^2 \sigma}{3\mu}  \sinh (2\mu |y|) \Bigg], \label{41}
\end{equation}
with $\mu=\sqrt{-\Lambda/6}$. And this brings us to the following
important conclusion: The Big Rip divergence on the brane, present
as we have seen when $\alpha >1/2$, becomes transferred into the
bulk. The bulk scale factor $a(t,y)$ diverges for arbitrary $y$ at
$t=t_s$, if $a_0(t)$ diverges at $t_s$. This result could hardly
have been seen beforehand, without calculation. There is moreover
no particular difference between an Einstein fluid and a modified
gravity fluid in this respect; they behave essentially in the same
way. It may also be of interest to note that if the brane is
tensionless, $\sigma=0$, then the bulk solution becomes quite
simple,
\begin{equation}
a^2(t,y)=\frac{1}{2}a_0^2(t)[1+\cosh(2\mu y)]. \label{42}
\end{equation}
The bulk scale factor thus increases exponentially on both sides
of the brane, for large $|y|$.

It would be of physical interest to understand the simultaneous
occurrence of singularities on the brane and in the bulk - perhaps
there are quantum effects at play here.

\bigskip



\end{document}